# Cycle-Consistent Adversarial Networks for Realistic Pervasive Change Generation in Remote Sensing Imagery

Christopher X. Ren, Amanda Ziemann,
James Theiler
Intelligence and Space Research Division
Los Alamos National Laboratory
Los Alamos, NM, USA
{cren, ziemann, jtheiler}@lanl.gov

Alice M.S. Durieux
Descartes Labs, Inc.
Santa Fe, NM, USA
alice@descarteslabs.com

*Abstract*—This paper introduces a new method of generating realistic pervasive changes in the context of evaluating the effectiveness of change detection algorithms in controlled settings. The method – a cycle-consistent adversarial network (CycleGAN) – requires low quantities of training data to generate realistic changes. Here we show an application of CycleGAN in creating realistic snow-covered scenes of multispectral Sentinel-2 imagery, and demonstrate how these images can be used as a test bed for anomalous change detection algorithms.

## I. INTRODUCTION

Change detection is a type of analysis in remote sensing in which we seek to identify differences between multiple co-registered images captured at different points in time [1], [2]. Algorithms developed to address this problem find applications in a variety of fields from disaster management (where they are applied to detect and delineate floods, droughts, and wildfires) [3] to monitoring urban areas [4], crops [5], and forests [6]. Within the field of change detection in remote sensing, we may further refine the question from *"What has changed?"* to *"Has a change occurred that is different from most other changes in the scene?"* Ziemann *et al.* [7] draw the analogy of two analysts: one is interested in seasonal variations resulting in pervasive changes such as drier vegetation, snowfall, *etc.*, whilst the other might only be interested in the construction of a new building in that same time period. Both of these types of changes result in spectral changes, thus presenting the challenge of translating the application and scale-dependent concept of "interesting" and "pervasive" changes to a mathematical framework that can be used to analyze remote sensing images. The detection of these "interesting" changes is the fundament of anomalous change detection (ACD) [8]–[10].

It is crucial to thoroughly test ACD algorithm performance on realistic data during the process of algorithm development. This can often be an issue due to the sparsity of data which contains both the anomalous change one might wish to detect, as well as strong pervasive changes which occur more broadly in the scene. Furthermore, even in the rare case where strong pervasive changes are present between available images (e.g., snowfall, drought, flooding, *etc.*), we are often only able to characterize ACD algorithm performance against one single type of pervasive change, as the same area may not experience two different types of pervasive change. To address this data sparsity issue, we utilize generative adversarial networks (GANs) [11] to generate realistic changes. GANs are a framework for estimating generative models, pitting a generative model against a discriminative model (adversary) whose objective is to classify whether a sample is from the generative model distribution or the data distribution [11]. A commonly used analogy paints the generative model as a forger attempting to produce falsified paintings, and the discriminative model as a detective attempting to differentiate between real paintings and forgeries.

## II. UNPAIRED IMAGE-TO-IMAGE TRANSLATION FRAMEWORK

### A. The Image-to-Image Translation Problem

In generating artificial pervasive changes we are seeking to "translate" an input (here, a remote sensing scene) to a corresponding output image (the same scene, but with generated pervasive changes). Just as a sentence can be translated from English to French, a scene may be translated from one domain to another; this is the concept of *image-to-image translation*. The potential of GANs as a general-purpose solution to image-to-image translation problems was first experimentally demonstrated by Isola *et al.* [12] in a paired-image setting. One of the key advancements GANs enabled in the image-to-image translation task is the learning of a *structured loss* which penalizes the overall configuration of the output as opposed to one in which each output pixel is considered independent from all others in the input image [12].

Zhu *et al.* developed the CycleGAN architecture [13] used in this work to learn from *unpaired* image sets. Given a set of images in domain $X$ and a different set in domain $Y$ we attempt to train a mapping $G : X \rightarrow Y$ such that the output $\hat{y} = G(x)$, $x \in X$ is indistinguishable from images $y \in Y$ by an adversary trained to differentiate between $\hat{y}$ and $y$, with the added requirement that learned translations should be "cycle consistent" such that if we learn two translators $G: X \rightarrow Y$ and $F: Y \rightarrow X$, then $G$ and $F$ should be inverses of one another and both translations should be bijections [13]. This cycle consistency loss ensures individual inputs and outputs are paired up in meaningful ways [13].

*B. CycleGAN Framework*

Here we describe the CycleGAN framework utilized in this work. Adversarial losses are applied to both mapping functions described above: $G$ and $F$. For $G$, which generates the translation $X \rightarrow Y$, and the discriminator $D_y$, which classifies samples as $\hat{y}$ (fake) or $y$ (real), the adversarial objective can be expressed as:

$$L_{GAN}(G, D_Y, X, Y) = \mathbb{E}_{y \sim p_{data}(y)}[\log D_Y(y)] + \mathbb{E}_{x \sim p_{data}(x)}[1 - \log D_Y G(x)]$$

And conversely the same objective is applied for the generator F which learns the mapping $Y \rightarrow X$.

The cycle consistency principle can be expressed as follows: for an image $x$ from domain $X$ we require $x \rightarrow G(x) \rightarrow F(G(x)) \approx x$ and conversely for an image $y$ from domain $Y: y \rightarrow F(y) \rightarrow G(F(y)) \approx y$. Thus, the cycle consistency loss is expressed as:

$$L_{cyc}(G, F) = \mathbb{E}_{x \sim p_{data}(x)}[||F(G(x)) - x||_1] + \mathbb{E}_{y \sim p_{data}(y)}[||G(F(y)) - y||_1]$$

The full objective is thus:

$$L(G, F, D_X, D_Y) = L_{GAN}(G, D_Y, X, Y) + L_{GAN}(F, D_X, X, Y) + \lambda L_{cyc}(G, F)$$

Here λ scales relative importance of the cycle consistency and adversarial losses. The generator/discriminator pairs $(G, D_Y)$ and $(F, D_X)$ are trained simultaneously in a pair of two-player min-max games linked by the cycle consistency loss as follows:

$$G^*, F^* = arg \min_{G,F} \max_{D_X, D_Y} L(G, F, D_X, D_Y)$$

Full details of the network architecture and training details used in this work can be found in [13].

### III. EXPERIMENTS

*A. Data Collection*

The translation we wish to apply as a pervasive change is transforming a non-snowy scene into a snow-covered one. To collect relevant data we utilized weather reports for areas across the USA and Europe to narrow down time ranges over which we were likely to observe snowfall in Sentinel-2 multispectral imagery; the time ranges chosen in this study were November 2018 to February 2019. Once the areas-of-interest (AOIs) for snowy scenes and approximate time ranges were established, we utilized the Descartes Labs platform [14] which provides streamlined access to publicly available satellite imagery from NASA and ESA, as well as the ability to filter imagery by cloud cover percentage. We use the same AOIs to collect images for our non-snow-covered scenes, but filtered over a time-period from May to August in 2018 and 2019.

Our final dataset consisted of 2154 non-snowy scenes and 1900 snowy scenes of size 600 by 600 pixels at a resolution of 10m per pixel, over 13 multispectral and derived bands. Of these, we reserve 200 of each domain for validation, although we note there is no strict validation loss in CycleGAN training.

*B. CycleGAN Training*

We train the CycleGAN architecture described in [13] with the overall objective described in Section II.B for 1000 epochs in total with an initial learning rate of 0.0002, decayed linearly after 500 epochs.

Fig. 1 shows collected Sentinel-2 images (top row), and their translations across domains (bottom row). We note the highly realistic artificially generated snow-covered scenes in the first column, demonstrating that it is indeed possible to use CycleGAN to produce realistic pervasive changes in remote sensing imagery.

*C. ACD on CycleGAN transformed Images*

When used in practice, ACD is applied to a pair of images that can be designated as a before/after pair, i.e., image $x$ and image $y$. As a baseline for comparison, we took a before/after pair of Sentinel-2 images and computed their ACD results. After completing the CycleGAN training, we apply the learned transformation on the after image to generate a new snowy image $\hat{y}$; we compare the ACD results for image x to image $\hat{y}$ against the baseline to show how these artificially generated pervasive changes can affect the results of an ACD algorithm. We focus on Hyperbolic Anomalous Change Detection (HACD) with local co-registration adjustment (LCRA) [15] here.

Fig. 2(a) and 2(b) show a pair of images taken over the Los Angeles Stadium at Hollywood Park. Fig. 2(c) shows the cycleGAN transformation applied to Fig. 2(b). Finally, we show how this transformation may affect the results of HACD with Fig. 2(d) showing HACD applied to the pair of real images, and Fig. 2(e) showing the results of HACD applied to the real initial image and the CycleGAN transformed image. We note that the construction of the stadium, which appears in the "Sentinel-2 to Sentinel-2" HACD detections, is masked by the simulated snowfall; only the shadows cast by the stadium are still detectable as anomalous changes after the CycleGAN transformation.

We repeat this experiment on another ROI, the Fagraskógarfjall landslide in West Iceland. Fig. 3(a) shows the untransformed image pair utilized to generate the ACD map shown in Fig. 3(b). The ACD map shows the region occupied by the landslide as a highly anomalous change, and also picks out the fact the landslide has blocked a river at the base of the mountains, causing an accumulation of water in the area. Figure 3(c) shows the image pair utilized to generate the ACD map in Figure 3(d). Here we see the cycleGAN-transformation hinders the ability of the HACD algorithm to pick out the landslide.

*D. Robust Detection Ratio Evaluation*

Finally, we quantify the "robustness" of our Sentinel-2 to Sentinel-2 HACD detections against the CycleGAN transformation by comparing the amount of overlapping detections for a given percentile threshold of the ACD values. To quantify this, we first apply HACD to the paired images, and output an array of ACD values for each pixel pair in the images. We then calculate a decision threshold as a percentile value of the ACD array. We use this decision threshold to select the ACD values above a given value as detections, and for a given threshold can evaluate the effect of the cycleGAN transformation on the ACD detections. To this end we use a "robust detection ratio" defined as:

$$\frac{X \cap Y}{X}$$

where X is the set of original Sentinel-2 to Sentinel-2 detections and Y is the new set of Sentinel-2 to CycleGAN detections. This is equivalent to asking the question: *For a given detection threshold, what ratio of the detected anomalous changes are robust to this particular cycleGAN transformation?*

Fig. 4 shows that even for a generous median threshold for ACD detections, only approximately 70% of the initial detections remain following the CycleGAN transformation. Past the 90$^{th}$ percentile we note the percentage of "robust" ACD detections drops drastically to below 30% for both scenes. This demonstrates the potential of this approach as a test bed for the robustness of various ACD algorithms, which may be applied to simulate desired pervasive changes.

## IV. CONCLUSIONS

We have shown in this work preliminary experiments which demonstrate the value of GANs in generating synthetic data to test remote sensing algorithms, in particular change detection algorithms. We note that whilst the cycleGAN framework shown here generates synthetic images of high perceptual quality, further work is needed to truly evaluate how the statistical nature of the image is altered by this transformation.

## ACKNOWLEDGMENT

This work was supported by the United States Department of Energy (DOE) through the Laboratory Directed Research and Development (LDRD) program at Los Alamos National Laboratory. **LA-UR-19-31936**.

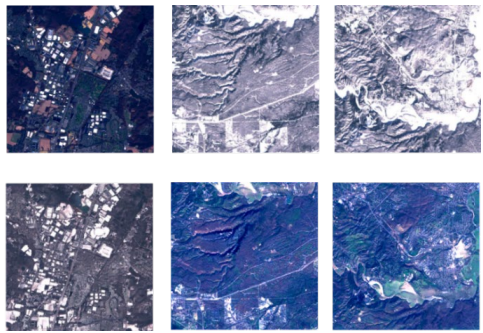

*Figure 1: Real Sentinel-2 images (top row) and their cycleGAN generated counterparts ( bottom row)*

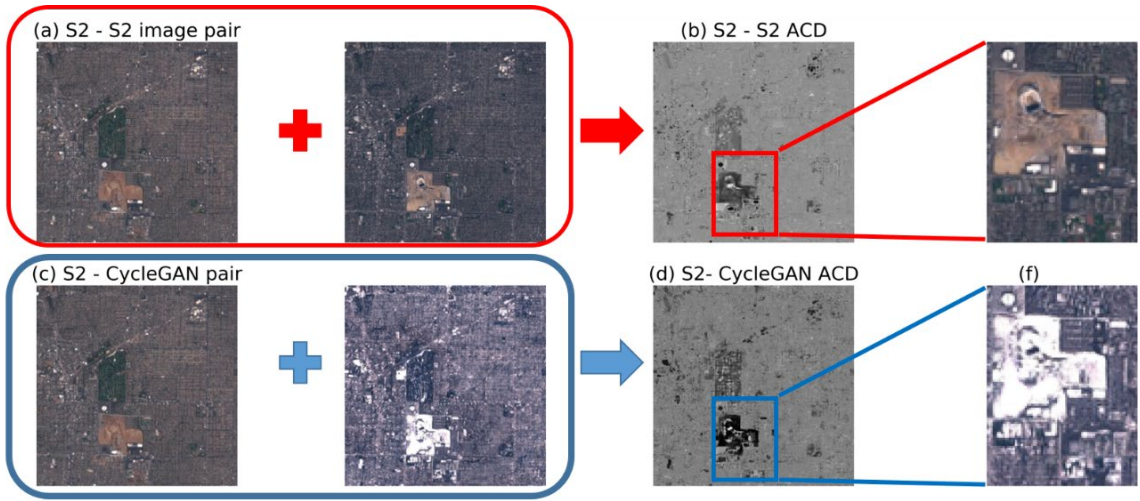

*Figure 2: (a) Sentinel-2 image pair used for ACD detections, taken over Los Angeles. (b) Grayscale HACD detection map (high values are bright), with an inset showing contribution of the stadium construction to the ACD values. (c) Sentinel-2 to CycleGAN transformed image pair used to produce the (d) ACD map.*

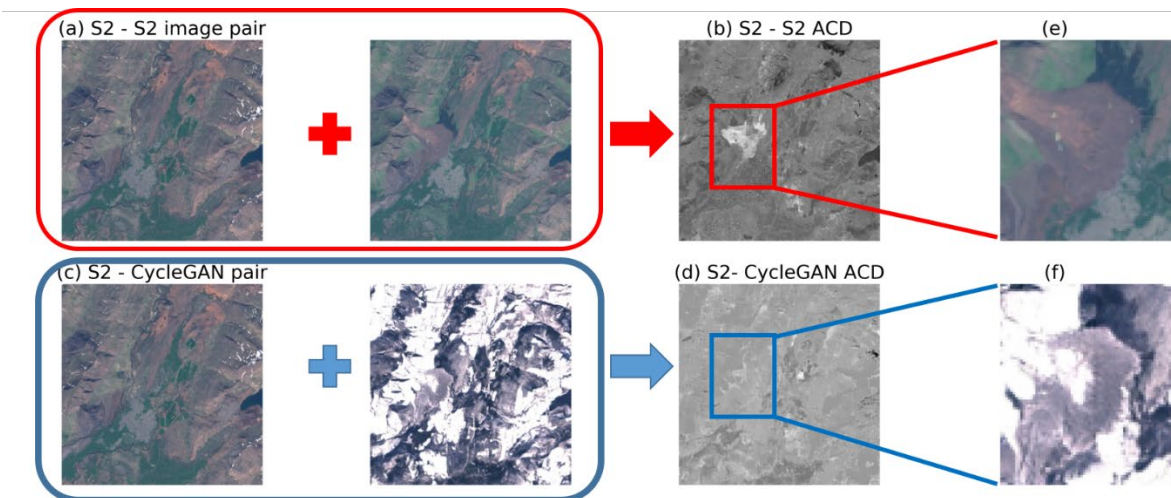

*Figure 3: (a) Sentinel-2 image pair used for ACD detections, taken over West Iceland. (b) Grayscale HACD detection map (high values are bright), with an inset showing contribution of the landslide to the ACD values. (c) Sentinel-2 to CycleGAN transformed image pair used to produce the (d) ACD map.*

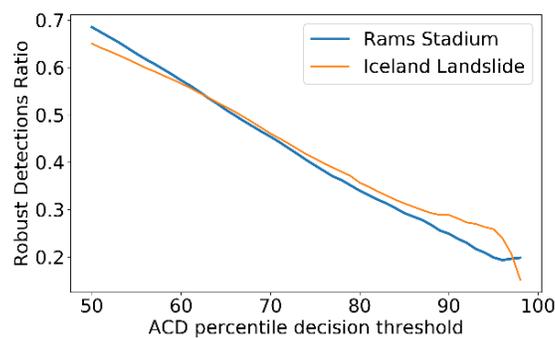

*Figure 3: Fraction of ACD detections that are 'robust' to the CycleGAN transformation, as a function of decision threshold.*